\documentclass[%
reprint,
superscriptaddress,
showpacs,preprintnumbers,
bibnotes,
 amsmath,amssymb,
 aps,
 prl,
floatfix,
]{revtex4-1}

\usepackage{setspace}
\usepackage{natbib}
\usepackage{graphicx}
\usepackage{dcolumn}
\usepackage{bm}
\usepackage[utf8]{inputenc}
\usepackage[T1]{fontenc}
\usepackage[english]{babel}
\usepackage{amsmath}
\usepackage{amsfonts}
\usepackage{amssymb}
\usepackage{commath}
\usepackage{makeidx}
\usepackage[caption=false]{subfig}
\usepackage{tikz}
\usepackage{soul}

\newlength\figureheight 
\newlength\figurewidth

\begin{document}

\title{Rigidity-controlled crossover: from  spinodal to critical failure}

\author{Hudson Borja da Rocha}
\email{hudson.borja-da-rocha@college-de-france.fr}
\affiliation{
LMS,  CNRS-UMR  7649,
Ecole Polytechnique, Université Paris-Saclay, 91128 Palaiseau,  France}
\affiliation{
 PMMH, CNRS-UMR 7636 PSL-ESPCI, 10 Rue Vauquelin, 75005 Paris, France}
\author{Lev Truskinovsky}%
\email{lev.truskinovsky@espci.fr}
 \affiliation{
 PMMH, CNRS-UMR 7636 PSL-ESPCI, 10 Rue Vauquelin, 75005 Paris, France}

\date{\today}
\begin{abstract}
Failure in disordered solids is accompanied by intermittent fluctuations extending over  a broad range of scales. The implied scaling has been previously associated with either spinodal or critical points.  We use an analytically transparent mean-field model to show that both analogies are relevant near the brittle-to-ductile transition. Our study indicates  that  in addition to the strength of quenched disorder, an appropriately chosen  global measure of rigidity (connectivity)  can be also used to tune the system to criticality. By interpreting rigidity as a timelike variable we reveal an intriguing parallel between earthquake-type critical failure and Burgers turbulence.
\end{abstract}
\maketitle

Failure in disordered solids takes place when elasticity (reversibility) breaks down  \cite{herrmann2014statistical}.  The  implied abrupt mechanical degradation can be  associated    with  brittle rupture \cite{Alava_2006}, large plastic avalanche \cite{Procaccia_PRE_2017}, or result from other nucleation type event \cite{Paco_PRL_2008}.  In strain controlled experiments,  failure may be accompanied by a dramatic stress drop, and the challenge is to predict and control such   undesirable events. 

The mechanism of failure in random elastic systems is nontrivial because of the intricate interplay between threshold-type nonlinearity, quenched disorder and long-range interactions. While the  strength  of disorder, the system size, and the range of elastic interactions are known to affect the failure mechanism \cite{Fisher_PRL_1997, Toussaint_PRE_2006, Sethna_PRL_2013, Roy_PRE_2017}, here we focus on the role of  system's \emph{rigidity}, which has recently emerged as another relevant factor in   failure-related phenomena \cite{Vitelli_PNAS_2016, Rocklin_PRM_2017, Nagel_NaturePhysics_2014}.

Failure in  disordered solids   is characterized by  scale-free statistics of large   events. The associated intermittency has been   linked to the existence of either spinodal    \citep{Selinger_1991_PhysRevA, Rundle_PRL_1989, Zapperi_PRL_1997, Wisitsorasak16068,Procaccia_PRE_2017} or critical points \citep{Moreno_PRL_2000, Sornette_PRL_1997,  Sethna_PRL_2013,Tarjus_PNAS_2018}.  At large disorder and infinite system size, failure is known to be linked to  percolation  \cite{Sahimi_PRL_1992, Roux_JSP_1988, Schmittbuhl_PRL_2003, Toussaint_PRE_2005, Andrade_PRL_2012};  however,  the physical nature of failure at finite disorder remains a subject of debate \cite{Sethna_PRL_2013, Tarjus_PNAS_2018, Wyart_PRE_2018, Procaccia_PRE_2017}.

In this Letter, we use an analytically tractable  mean-field model to show that both spinodal and critical scaling behaviors can coexist near the threshold of the brittle-to-ductile transition \cite{Herrmann_PRB_1988, Huang_2014, Liu_2019, Eckert_2016, SUBHASH_2006, ZHAO_2018, SELEZNEVA_2018}.  Ductile response is understood here in the sense of stable development of small avalanches  representing micro-failure events \cite{Krajcinovic1998, Gao_PRSA_2018}. Brittle response  necessarily involves  large  events representing   system size instabilities \cite{papanikolaou2017brittle, berthier2018rigidity}. 
      
Our starting point is the fiber bundle model (FBM) with global stress redistribution \citep{hansen2015fiber}.  This model was used to explain a variety of physical phenomena from  failure   of textiles  \cite{Peirce}, and   acoustic emission  in loaded composites \cite{Nechad_JMPS_2005} to  earthquake dynamics \cite{Sornette_1992}.  It is usually studied in the \emph{stress} control setting, where failure is brittle and scaling is spinodal \citep{hansen_1992, Alava_2006}. To address failure under \emph{strain} control  and to be able to tune the system to criticality,  we drive the system differently,  using an external harmonic  spring  \citep{Roux_IJSS_1999, Paco_PRL_2008}.

In our analysis, brittle failure emerges as a supercritical, while  ductile failure as  a subcritical phenomenon.  The critical behavior can be  associated with the brittle-to-ductile transition and  we show that due to superuniversality of the mean-field models \cite{Tarjus_PRB_2014}, the equilibrium and out-of-equilibrium exponents are the same.

The main focus of this Letter, however,  is the role of the system's rigidity \cite{Manning_PNAS_2019}  as the regulator of the brittle-to-ductile transition. It is known that rigid,  crystal-like solids subjected to stresses fail catastrophically  \cite{Rice_1968}.  Instead loose, marginally jammed solids fail gradually \cite{Vitelli_PNAS_2016, Rocklin_PRM_2017, Nagel_NaturePhysics_2014}. In view of the minimal nature of our model, we could construct analytically the  rigidity-disorder phase diagram delineating the domain of ductile behavior at low rigidity and high disorder from the domain of brittle behavior at low disorder and high rigidity. 

One of our crucial findings is that in the brittle-to-ductile crossover region, which bridges \emph{robust} spinodal criticality with \emph{tuned} classical criticality,  the  transitional  exponents are  non-universal, depending sensitively on system size,  disorder, and rigidity. We also show that when rigidity can be conditioned by the system size, failure becomes brittle in the thermodynamic limit, and scaling survives only as a finite size effect, cf. \cite{Sethna_PRL_2013, Vitelli_PNAS_2016}.
 
Equilibrium (static) avalanches, corresponding to jumps between different globally minimizing configurations,  have been previously linked to Burgers shocks  \cite{Bouchaud_1997, Le_Doussal_PRE_2009}. Here we extend this analogy showing that if rigidity is interpreted  as "time", and strain as "space", the brittle-to-ductile transition and the associated critical behavior can be viewed as a  "finite time" Burgers  turbulence  \cite{BEC20071}.  Given that our model is essentially a mean-field version of the  Burridge-Knopoff model  \cite{Sornette_1992}, the developed analogy reinforces  a  conceptual link  between  earthquakes (fracture) and  turbulence \cite{Basu2019}.

\begin{figure}[ht]
\centering
\includegraphics[width=0.48\textwidth]{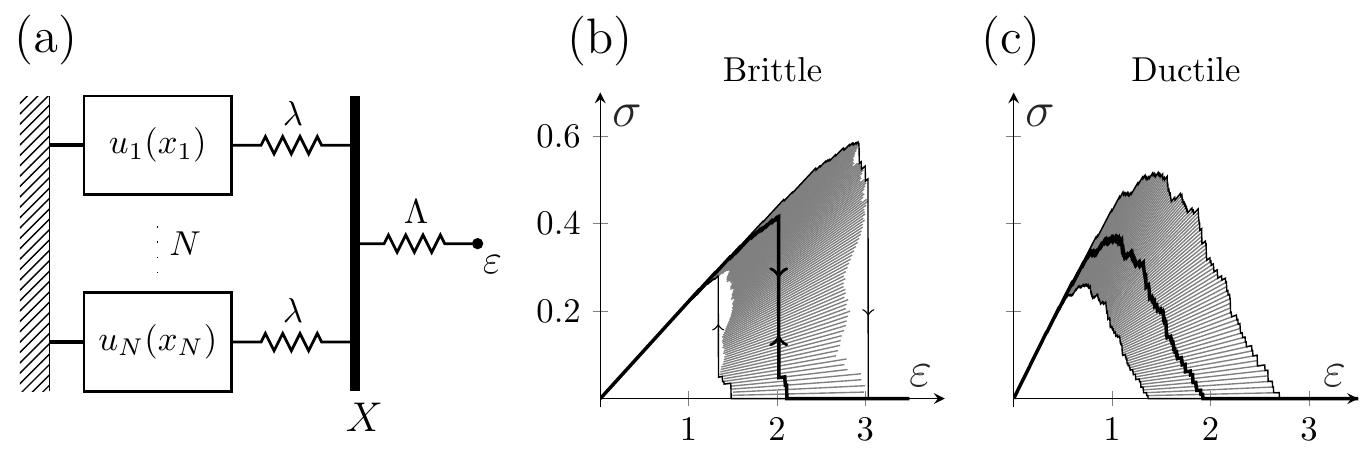}
 \caption{(a) Schematic representation of the system; (b) brittle reponse at $\Lambda=0.4$; (c) ductile response at $\Lambda=5$.  Solid black lines: equilibrium path, thin black line: out-of-equilibrium paths; grey lines: metastable states. Parameters: $N=100$, $\lambda=1$, $\rho=4$.  }\label{fig:sigma}
\end{figure}

Consider a discrete system with dimensionless energy: 
\begin{equation}\label{eq:Hamiltonian}
\mathcal{H}={\displaystyle \frac{1}{N}\sum_{i=1}^{N}}\left[u_i(x_{i})+\frac{\lambda}{2}\left(X -x_{i}\right)^{2}\right]+\frac{\Lambda}{2}( \varepsilon-X)^{2},
\end{equation}
where $u_i(x)$ is a Lennard-Jones type potential of a breakable element, $X$ is a Weiss-type mean field accounting for  the  interaction among   breakable   elements, and   $\varepsilon$ is the controlling parameter representing the harmonic interaction of the field $X$ with the loading  device, see Fig.~ \ref{fig:sigma}(a).  For determinacy, we assume that the  potential $u_i(x)$ is piece-wise quadratic: 
 $u_i(x)=(x ^{2}/2)  \Theta (l_i-x)+(l_i^{2}/2)  \Theta (x-l_i)$, where  $ \Theta  $ is the Heaviside function;  for $x\leq l_i$, the element is intact, while for $x>l_i$, it  is broken. Here,  $l_i$ are random  numbers drawn from the  probability distribution $f(l)$.   In our  numerical illustrations, we use   Weibull's  distribution with density $f(x)= \rho x^{\rho-1}\exp{(-x^\rho)}$;   broad disorder corresponds  to small $\rho$  \footnote{If the breakable elements are  composed of sub-parts linked in series and  if the failure is associated with breaking of the weakest sub-part, the Weibull distribution emerges rigorously  in the thermodynamic limit  as the  distribution for the breaking threshold of the whole system. Here we assume  that  the distribution of thresholds for  the  sub-parts has a compact support. }.  In our generalization of the FBM \eqref{eq:Hamiltonian}, we  introduced two new parameters: the  internal stiffness $\lambda $, and the external stiffness $\Lambda = \kappa/ N$,  where $\kappa \sim N$ is the   effective  elasticity of the elastic environment \citep{Roux_IJSS_1999}. 
  
In Fig.~\ref{fig:sigma}(b,c),  we illustrate the typical behavior of the   local and global minima  of \eqref{eq:Hamiltonian} by showing the  relation between the applied strain $\varepsilon$ and the conjugate stress $\sigma=\Lambda(\varepsilon-X)$, see also \cite{SOM}.   Our Fig.~\ref{fig:sigma}(b) shows  the brittle behavior, which includes a  system size transition from the partially broken  to the fully broken state.  In  contrast, our   Fig. \ref{fig:sigma}(c)  illustrates the ductile behavior, characterized by the  gradual accumulation of damage. The \emph{equilibrium} (global minimum) deformation paths are shown in Fig. \ref{fig:sigma}(b,c)  by thick black lines. We assume that failure is reversible, and show by thin black lines the\emph{ out-of-equilibrium} (marginally stable) paths that are different for loading and unloading. 
  
The   boundary separating  brittle and ductile regimes  depends  on the strength  of the disorder (our parameter $\rho$) and on the dimensionless  parameter 
\begin{equation}
\nu=\frac{\lambda}{\Lambda(1+\lambda)},
\end{equation}
which we interpret as a measure of the structural  \emph{rigidity} of the system \cite{Vermeulen_PRE_2017, crapo1979structural, Strogatz_Nature_2019, Manning_PNAS_2019}. When $\nu$ is small, meaning that  either $\Lambda$ is large or $\lambda$  is small,  individual breakable elements interact weakly and  the limit $\lambda  \to 0$ can be   associated with  the  (jamming) threshold beyond which the rigidity is lost \cite{Nagel_NaturePhysics_2014}. Instead, when $\nu \to \infty$  the system can be viewed as overconstrained \cite{Vitelli_PNAS_2016,Rocklin_PRM_2017}.
 
\begin{figure}[t]
\centering
\includegraphics[width=.48 \textwidth]{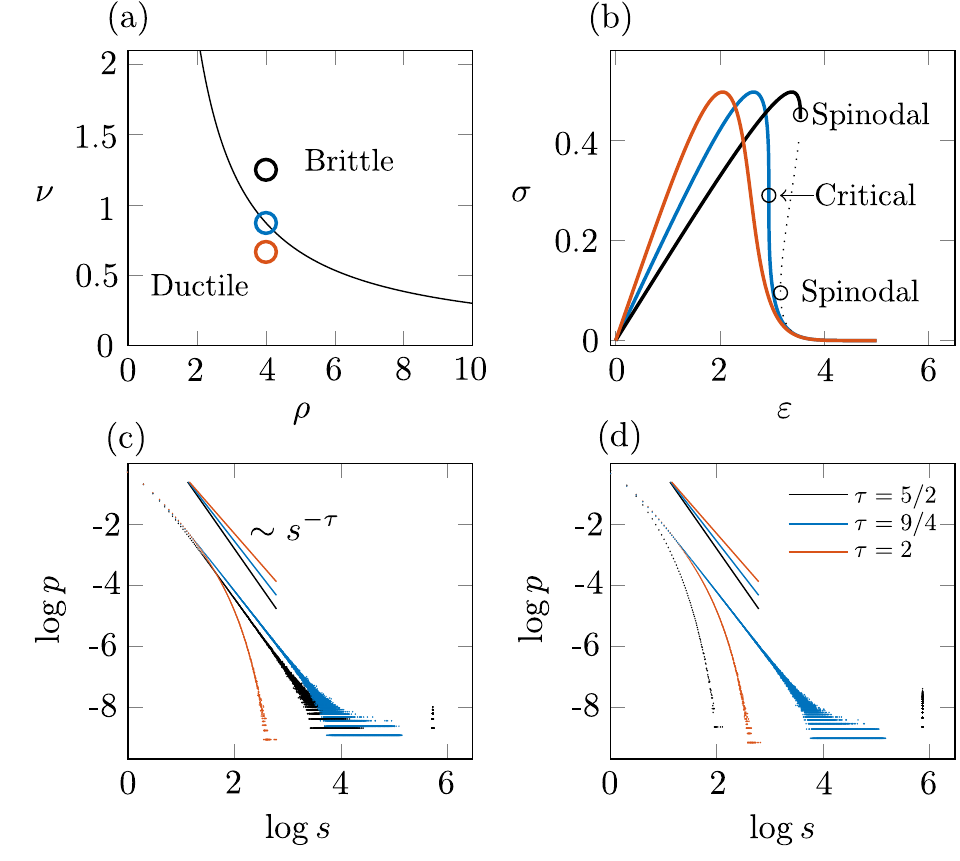}
 \caption{(a) Ensemble averaged brittle-to-ductile transition  line;  (b) typical averaged stress-strain curves; (c) non-equilibrium  avalanche distribution; (d)    equilibrium avalanche distribution.  Parameters:  $N= 10^6$, $\lambda=1$. The avalanche distributions was averaged over $10^4$ realizations.}
\label{fig:popsnap}
\end{figure}
 
The ensemble averaged brittleness/ductility threshold can be found by solving  the system of equations $-2 f(x_c) =f'(x_c) x_c$ and  $ 1-F(x_c) =f(x_c)x_c-1/\nu$,  where  $F(x)=\int_{0}^{x}f(x')dx'$  is  the cumulative distribution of thresholds \cite{SOM}.  In particular, for Weibull-distributed thresholds, the line separating brittle from ductile behavior is given by the equation $\nu_*=\exp{(1/\rho +1)} /\rho$, see  Fig. \ref{fig:popsnap}(a).  

In the  limit $N \to \infty$ the avalanche distribution   in the model \eqref{eq:Hamiltonian} can be computed analytically \cite{hansen2015fiber,SOM}
\begin{equation}
p(s)= \frac{s^{s-1}}{s!}  \int_0^{\infty} \frac{\left(1-g(x)\right) f(x)}{g(x)}e^{-h(x)s} dx,
\end{equation} 
where  $g(x)=  f(x)x/(1-F(x)+\nu^{-1})$, and $ h(x)=g(x)-\ln g(x)$. In  the large-event-size asymptotics the universal pre-integral multiplier $s^{s-1}/s!\sim s^{-3/2}$   represents the  classical  mean-field  contribution,   reflecting the built-in statistics of Brownian return times \cite{Sornette_1992, Zapperi_PRL_1995}. In the limit $s \to \infty$   the  integrated  distribution can be obtained by   the   saddle-point approximation around  the global minimum,  $x_0$, of the function $ h(x)$  \footnote{Our  asymptotic analysis  is  valid for arbitrary disorder as long as the function $h(x)$ has a minimum.  Some    long tailed disorders  can modify the behavior of the system, for instance,  the distribution of thresholds  $F(x)=0 $, for $ x\leq 1$, and $F(x)=1-1/\sqrt{x}$, for  $x>1$,  leads to the function  $h(x)$ without a   minimum.}.  It is a root of  either $g(x_0)=1$ or $g'(x_0)=0$, and the emergence  of such two  cases   is  a general feature of  mean-field models \cite{Munoz_PRL_2016}.

Consider first the out-of-equilibrium path (dynamic avalanches).  Then, if
  $g'(x_0)=0$ while  $g(x_0)\neq 1$,  we obtain $p(s) \sim s^{-2} e^{-s(h(x_0)-1)}$. This is a sub-critical distribution describing the ductile (POP) regime  \cite{footnote},   dominated by uncorrelated random events.   If  $g(x_0)=1$  but  $g'(x_0)\neq 0$,   the point $x_0$ is  spinodal, and the distribution is   super-critical, characterizing the brittle (SNAP) regime  \cite{footnote}. Neglecting the system-size   events,  we can  write the corresponding local distribution in the form  
$
p(s,x)\sim s^{-3/2}(x-x_0)e^{-s\frac{g'(x_0)^2}{2}(x-x_0)^2}.
$
The avalanche size diverges near  $x_0$, and the integrated distribution takes the classical form $p(s)\sim s^{-5/2}$ \citep{hansen2015fiber}.  Finally, if $g(x_0)=1$ and $g'(x_0)=0$,  the local distribution reads
$
p(s,x)\sim s^{-3/2}(x-x_0)^2e^{-s\frac{g''(x_0)^2}{4!}(x-x_0)^4}.
$
The characteristic avalanche size   again diverges near   $x_0$  and the integrated  distribution takes the form
$
p(s)\sim s^{-9/4}.
$
This is the  critical  (crackling) regime  \cite{footnote}  associated with brittle-to-ductile transition;  the  exponent $9/4$ has appeared previously in the  context  of  composite FBM  involving   breakable and unbreakable  springs \cite{Kun_EPL_2008}.   Other values of the exponents also appeared  in the more complex FBM based models describing richer physics  \cite{Hidalgo_PRE_2009}.

The computed critical exponents   coincide  with  the ones known for the mean-field RFIM   \citep{Dahmen_PRB_1996, Zapperi_PRL_1997}, because   the energy \eqref{eq:Hamiltonian}  can be mapped on  the  soft-spin RFIM. To this end we need to  minimize out   the variable $X$, which gives 
$\mathcal{H}=-(1/N^2)\sum_{i,j}Jx_i x_j-(1/N)\sum_i[Hx_i-v_i(x_i)],$
where $v_i(x)=u_i(x)+x^2 +  \lambda\Lambda\varepsilon / 2(\lambda+\Lambda)$, see also \cite{SOM}.  Note that   the Lennard-Jones type potential $u_i(x)$ was transformed along the way into the \emph{double-well } potential $v_i(x)$. Other  mean-field formulations  leading to the same spinodal and critical exponents  that are relevant for amorphous plasticity are  discussed in \cite{Tarjus_PNAS_2018,Wyart_PRE_2018}; the same two main  regimes have been also identified  for some  sandpile automata \cite{Munoz_PRL_2016}. Interestingly, a numerical analysis  of a  \emph{non-mean-field} model of   a structural  phase transition reveals the possibility of a similar  coexistence of   two   scaling  behaviors \cite{Paco_PRL_2008}. 
\begin{figure}[ht]
\centering
\includegraphics[scale=1]{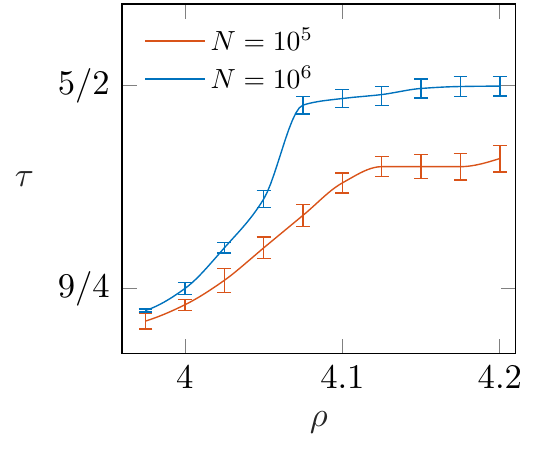}
 \caption{Finite size  crossover associated with  brittle to ductile transition.  Each  curve  gives the  value of the scaling exponent  averaged over 250 realizations. Parameters $\lambda=1$ and $\nu =0.873$ (critical value at $\rho=4$).  The exponents and the  uncertainty were computed  using the maximum likelihood method \cite{clauset_2009}. }
\label{fig:ExponentVar}
\end{figure}
 
In  finite size systems, the crossover from the \emph{robust} spinodal scaling in the brittle regime (exponent $5/2$)  to the \emph{non-robust} critical scaling (exponent $9/4$)  takes place in an extended  transition zone, where  the system exhibits non-universal  exponents, see  Fig.~\ref{fig:ExponentVar}.  The ubiquity of such transitional  phenomena may  explain  the large scatter in reported scaling behavior of disordered solids \cite{LT_PRL_2015, Vives_PRE_2019, Sparks_PRM_2018}.

The  mean-field model \eqref{eq:Hamiltonian} can be used to demonstrate directly  the   \emph{super-universality} of the critical regime \citep{Tarjus_PRB_2014, Vives_PRB_2004, Maritan_PRL_1994, Dahmen_PRE_2009, Tarjus_PRB_2018}. For instance, one can show  that  the exponent $9/4$  is valid for both out-of-equilibrium   and equilibrium  paths \cite{SOM}. Instead, the spinodal criticality, which exists in the out-of-equilibrium model, disappears   in the equilibrium model  because the  SNAP event  takes place   before the spinodal point is reached. Integrating  the avalanches should be then performed only up to  some  Maxwellian $x_*<x_0 $, and since in this case  the function $h(x)$  attains its minimum at the boundary,  we obtain
$p(s) \propto s^{-5/2}e^{-s(1-h(x_*))}.$
While this distribution has the same exponent $5/2$ as in the case of the out-of-equilibrium path, the  scaling is now  obscured   by the  exponential cut off. 

\begin{figure}[ht]
\centering
\includegraphics[scale=.9]{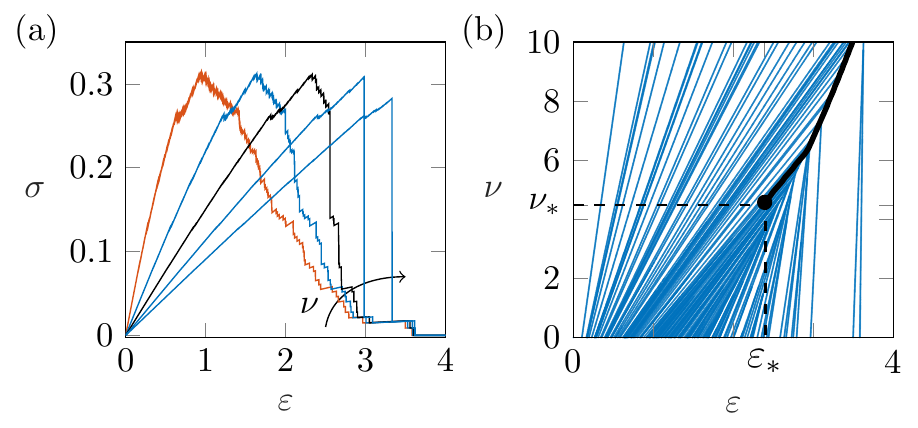}
 \caption{(a) Time (rigidity) evolution of the randomly distributed initial Burgers data $\sigma_0(\varepsilon)$ (red) at $\rho=2$, and $\lambda=1$;   black line corresponds to  $\nu_*=\exp{(1/\rho +1)} /\rho$.   (b) Shock merging with critical complexity appearing at  $\nu=\nu_*$.  Thick black line shows the shock in  the averaged system  which emerges in  the limit  $N \to \infty$. }
\label{fig:random_shock}
\end{figure} 
 
We now turn to   an intriguing analogy  between the  equilibrium version of the model \eqref{eq:Hamiltonian} and  Burgers turbulence  \cite{BEC20071}.  If we minimize out  the  variables  $x_i$ in \eqref{eq:Hamiltonian} and consider the thermodynamic limit  $N \to \infty$ \cite{SOM}, the equilibrium problem reduces to finding $\tilde{\mathcal{H}}(\varepsilon,\nu) \sim \min_{X\in\mathbb{R}}\left\lbrace \frac{1}{2 \nu}(\varepsilon-X)^2+q^{\infty}(X)\right\rbrace,$
where 
 $q^{\infty}(z)=[1-F(\sqrt{\lambda/(\lambda+1)} z)](z^2/2)+\sqrt{\lambda/(\lambda+1)}\int_{0}^z f(\sqrt{\lambda/(\lambda+1)}z')(z'^2/2)dz'.$  We can now use the Hopf-Lax formula \cite{Evans_PDE} to  turn this variational problem into a Cauchy problem for a  Hamilton-Jacobi equation  $ \partial_\nu\mathcal{\tilde H}  +\frac{1}{2}(\partial_\varepsilon\mathcal{\tilde H})^2 = 0,$ where  the rigidity $\nu$ plays   the role of time.   This equation must be supplemented by the  initial condition  $\mathcal{\tilde H}(\varepsilon,0) = q^{\infty}(\varepsilon)$. Then the  tension   $\sigma=\partial_\varepsilon\mathcal{\tilde H}$ satisfies the inviscid Burgers equation
 \begin{equation} \label{eq:burgers}
  \partial_{\nu} \sigma+\sigma\partial_\varepsilon\sigma = 0
  \end{equation}
with  initial condition $\sigma_0=\partial_\varepsilon q^{\infty}(\varepsilon)$. Interestingly, the viscous Burgers equation for $\sigma$ and the corresponding KPZ equation \cite{KPZ_PRL_1986}  for  $\mathcal{H}$ emerge as a finite size effect in the model  \eqref{eq:Hamiltonian}  with  finite temperature.
 
As a result of the   reduction of the problem \eqref{eq:Hamiltonian}  to \eqref{eq:burgers},  avalanches become   shock waves  \cite{Bouchaud_1997}.  In the  averaged model,   the    ductile-to-brittle transition   can be then associated with the shock formation  at  a finite rigidity  $\nu_*=\min_{\varepsilon\in \mathbb{R}}\left\lbrace - 1/\partial_\varepsilon\sigma_0(\varepsilon)\right\rbrace$,  see Fig. \ref{fig:random_shock}(b); in the $(\varepsilon,\nu)$  plane  this "event"  becomes  a  direct analog of  the   liquid-vapor critical point.   
 
At finite $N$,  the "evolution" equation for the stress remains the same as in the case $N \to \infty$, while   the initial condition   changes  to $\sigma_0=\partial_\varepsilon q(\varepsilon)= N^{-1}\sum_{i=i}^N \varepsilon \Theta (l_i-\sqrt{[\lambda/(\lambda+1)]}\varepsilon)$, see   \cite{SOM} for details.  In Fig.~\ref{fig:random_shock}(a)  we show how the increase of rigidity   transforms the ductile response,  where    avalanches   take  the form of   small Burgers shocks (POP events),  into the brittle response with a single Burgers shock representing a  system size SNAP event.  In Fig.~\ref{fig:random_shock}(b) we track   the position of individual shocks and  visualize their merging  sequence. 
 
\begin{figure}[ht]
\centering
\includegraphics[scale=1]{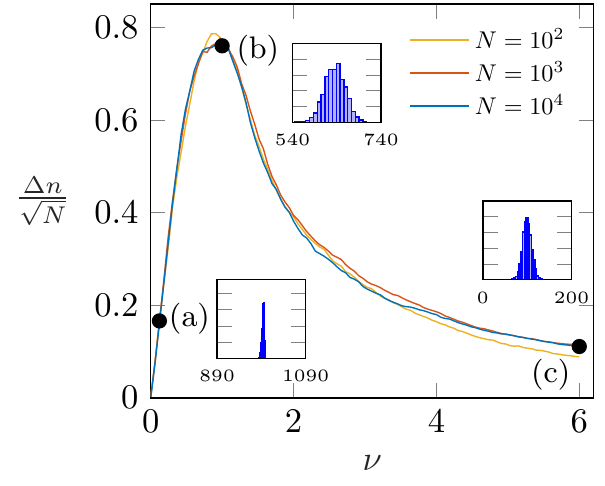}
\caption{Time (rigidity) evolution of the normalized standard deviation for the number of shocks. The statistics was obtained from  $K=1000$ realizations of the quenched disorder  with $\rho=3$, and $\lambda=1$. Inset plots: (a) $\nu= 0.1$ (b) $\nu=1$; (c) $\nu= 6$. } \label{fig:normalized_mean_std}
\end{figure}

To highlight the \emph{critical} nature of the system  with  rigidity value close to $\nu_*$, we  studied  the  $\nu$  dependence of the number of shocks $n$.  In Fig.~\ref{fig:normalized_mean_std}, we show the  standard deviation $\Delta n=[ K^{-1}\sum_{i=1}^{K} (n_i- K^{-1}\sum_{i=1}^{K} n_i)^2]^{1/2}$, where  different realizations of  disorder are indexed by $i=1,2,..., K$.   Note the  peak indicating   the anomalous broadening  of the distribution around  the critical point  $ \nu=\nu_*$.   The situation  is fundamentally different in  the conventional decaying  Burgers turbulence where the initial data   have  \emph{zero} average, which infinitely delays  the emergence  of scaling.  
 
So far we were assuming that   the rigidity measure  $\nu $ is finite  as  $N \to \infty$.   A broader class of  elastic environments  can be modeled if we assume   that  $\kappa \sim N^\alpha$, with $0\leq \alpha\leq 1$.  For instance,  if  the   load   is transmitted through  a surface  of  a 3D body   we have   $\alpha=2/3$ and $\nu \sim  N^{1/3}$.  In this setting,   small systems  would be necessarily ductile, while brittle behavior would dominate in the thermodynamic limit.  At a given disorder, the scaling  will be then seen    in  a window of system sizes,  while the (percolation type) critical regime  will emerge only at   infinite size  and infinite disorder   \citep{Olami_PRL_1992,Toussaint_PRE_2006,  Sethna_PRL_2013}. 
 
To conclude,  we used an analytically transparent model to quantify the role of system's rigidity (global connectivity) as a  control parameter for the transition from brittle to ductile failure. We showed that this transition can be associated with the crossover from spinodal to classical criticality,  generating,  in finite size systems, a  scaling region with non-universal exponents.   Such behavior is  generic for a broad class of systems, encompassing fracture, plasticity,  structural phase transitions, and now we established  a new  link to fluid  turbulence. 
 
\begin{acknowledgments}
The authors are grateful to R. Garcia-Garcia, K. Dahmen, and M. Mungan for helpful discussions.   H.B.R. was supported by a PhD fellowship from Ecole Polytechnique; L. T. was supported by the grant  ANR-10-IDEX-0001-02 PSL.
\end{acknowledgments}
%

%

\section{Supplemental Material }

To obtain  the avalanche distribution in our generalized FBM problem with controlled  \emph{length}, we follow the general methodology largely developed by Hansen and collaborators in their studies of the classical FBM  problem which implies control of  the \emph{force} \citep{hansen2015fiberSM, Hansen:262741SM, hansen_1992SM, KlosterSM, Pradhan:2010aaSM}. 

\paragraph{Metastable states.}
First, we use the condition  $\partial_{X}\mathcal{H}=0$ to obtain
$
X(\boldsymbol{x},\varepsilon)={\frac{1}{\lambda+\Lambda}{ \left(\Lambda\varepsilon+\lambda\frac{1}{N}\sum_{i=1}^{N}x_{i}\right)}},
$
and the condition   $\partial_{x_i}\mathcal{H}=0$ to obtain  $u'(x_i)=\lambda(X-x_{i})$. In view of permutational invariance, we can characterize the microscopic state  by the number of broken bonds, $k$, 
which  gives 
\begin{equation}\label{eq:equiy}
\hat{X}(k,\varepsilon)={\displaystyle \frac{(1+\lambda)\Lambda\varepsilon}{\lambda (1-k/N)+\lambda \Lambda +\Lambda}}.
\end{equation}
For the attached links  we have 
\begin{equation}\label{eq:x0}
\hat{x}_{0}(k,\varepsilon)={\displaystyle \frac{\lambda\Lambda\varepsilon}{\lambda (1-k/N)+\lambda \Lambda +\Lambda}},
\end{equation}
and for the broken links 
\begin{equation}\label{eq:x1}
\hat{x}_{1}(k,\varepsilon)={\displaystyle \frac{(1+\lambda)\Lambda\varepsilon}{\lambda (1-k/N)+\lambda \Lambda +\Lambda}}.
\end{equation}
The energy of  the equilibrium configurations can be written as
\begin{equation}\label{eq:enehd}
\mathcal{H}(k,\varepsilon)=a_k \varepsilon^2+S_k,
\end{equation}
where $a_k=\displaystyle \frac{1}{2}\frac{\lambda \Lambda(N-k)}{\lambda(N-k) + N (\lambda \Lambda + \Lambda)}$, and  $S_k$ is the energy of the broken  bonds. If  $\bar{x}_i, i=1,...,N$ is the ordered sequence of failure thresholds,  $\bar{x}_1\leq \bar{x}_2\leq \dots\leq \bar{x}_N $, we can write   $S_k=\displaystyle\frac{1}{N}  \sum_{i=1}^{k}\frac{\bar{x}_i^2}{2}$, and   $S_0=0$.  We observe  that  $a_k$ is a (strictly) monotonically decreasing sequence while $S_k$ is a (strictly) monotonically increasing sequence.  

The stress-strain relation for a microscopic state characterized by the parameter $k$ is
\begin{equation}\label{eq:tensionhd}
\sigma(k,\varepsilon)=\frac{\partial\mathcal{H}(k,\varepsilon)}{\partial \varepsilon}=\displaystyle  \frac{\lambda \Lambda(N-k)\varepsilon}{\lambda(N-k) + N (\lambda \Lambda + \Lambda)}.
\end{equation}
Each value of $k$ defines an equilibrium branch extending between the two limits  induced by the inequalities $\hat{x}_{0}(k,\varepsilon)<\bar{x}_k$   and $\hat{x}_{1}(k,\varepsilon)> \bar{x}_k$. For the failure thresholds we can then write
\begin{equation}\label{eq:zsup}
{\displaystyle \varepsilon^{f}_k=\frac{\lambda+1}{\lambda }\left[\left(1-\frac{k}{N}\right)\nu + 1 \right]\bar{x}_k}, 
\end{equation}
where $ 0\leq k<N$. Similar expressions can be obtained for the  rebuilding thresholds 
\begin{equation}\label{eq:zk_breakingnf}
{\displaystyle \varepsilon^{r}_k= \left[\left(1-\frac{k}{N}\right)\nu+1 \right]\bar{x}_k}, 
\end{equation}
where  $ 0< k\leq N$. The ensuing equilibrium branches are represented by the gray lines in Fig.~1 (b, c) in the main text. 
 
To analyze their (local) stability, we need to study the positive definiteness of the Hessian matrix for the energy   
$\mathcal{H}(\boldsymbol{x},X)$ 
\begin{equation}
  \boldsymbol{ \mathcal{M}}=
\begin{pmatrix}
M_1 & 0 & \dots & 0 & -\lambda \\ 
0 & \ddots & \ddots & \vdots & \vdots \\ 
\vdots & \ddots & \ddots & 0 & \vdots \\ 
0 & \dots & 0 & M_N & -\lambda \\ 
-\lambda & \dots & \dots & -\lambda & N(\lambda+\Lambda)
\end{pmatrix} ,
\end{equation}
where $M_i$ is either $\lambda+1$, for $1\leq i < N-k$, or $\lambda$, for $N-k \leq i \leq N$. The sufficient condition for   stability is that all the principal minors of $\mathcal{M}$ are positive.  The first $N$ minors are just the product of diagonal therms  are therefore always positive. The last principal minor, the determinant
\begin{equation}
\det(\mathcal{ \boldsymbol{ \mathcal{M}}})=\prod_{i=1}^N M_i \sum_{i=1}^{N} \left(\lambda+\Lambda- \frac{\lambda^2}{M_i}\right).
\end{equation}
 is also positive  implying  stability of the obtained equilibrium configurations; the unstable configurations must contain at least one element in the spinodal state represented in our  model by a single point.

\paragraph{Equilibrium (global minimum) path.} For large  $N$ we can write 
\begin{equation}
S_k=\frac{1}{N}\sum_{i=0}^{k} \frac{\bar{x}_i^2}{2}\approx \int_{\bar{x}_{1}}^{\bar{x}_k}  \frac{x^2}{2}f(x)dx,
\end{equation}
where we used the fact   that for ordered distributions we can use the approximation  $k/N \sim F(\bar{x}_k)$ \cite{arnold1992firstSM}.  We can then  write the  continuous approximation of the discrete energy in the form
\begin{equation}
\mathcal{H}(x,\varepsilon)=\frac{\lambda\Lambda (1-F(x))}{\lambda(1-F(x))+\Lambda(\lambda+1)}\frac{\varepsilon^2}{2}+\int_{0}^{x} f(x')\frac{x'^2}{2}dx'.
\end{equation}
Using the equilibrium condition $\partial \mathcal{H}(x,\varepsilon)/\partial x =0$, and applying it  for  the  discrete values  $\bar{x}_k$, we obtain
\begin{equation}\label{eq:zk_g}
\varepsilon^g_k=\frac{1}{\sqrt{\Lambda \nu}}\left[\left(1-\frac{k}{N}\right)\nu+1\right]\bar{x}_k.
\end{equation}
Note that the  three formulas (\ref{eq:zsup}), (\ref{eq:zk_breakingnf}) and  (\ref{eq:zk_g}) are different only by constant multipliers.
 
\begin{figure}[ht]
\centering
\includegraphics[width=.45\textwidth]{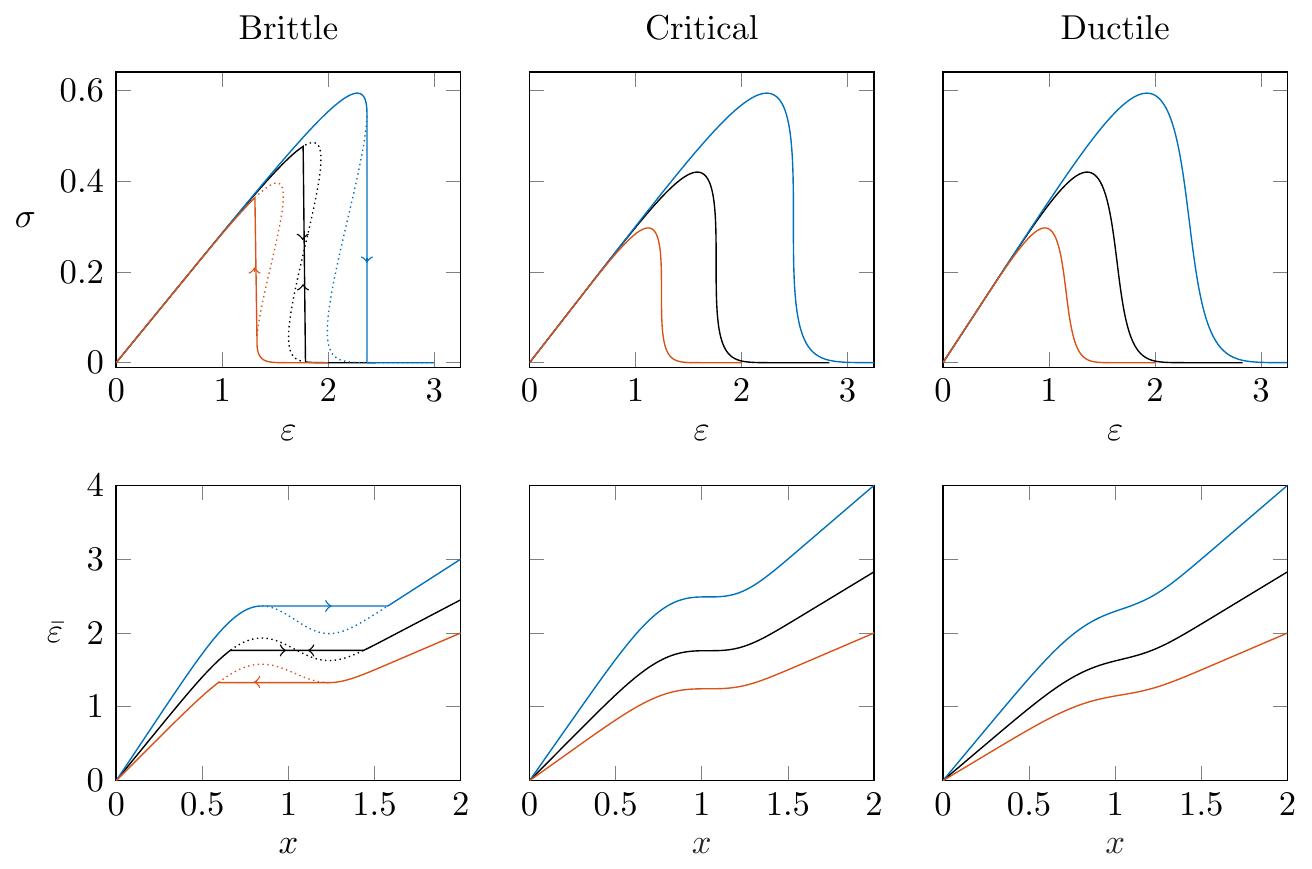}
\caption{First row: avergaed stress-strain relations , second row: averaged strain dependence on  the  internal variable $x$; blue (red) curves correspond to the loading (unloading) out-of-equilibrium paths;  black curves correspond to the  equilibrium (global minimum) path.} \label{fig:Average_properties}
\end{figure}

\paragraph{Out-of-equilibrium (zero viscosity limit) path.} Each microscopic configuration characterized by parameter $k$ exists in an extended domain of the loading parameter $\varepsilon$ between the failure strain
$\varepsilon_k^f$ and the rebinding strain 
$
\varepsilon_k^r$. At large $N$, we can use the approximation 
\begin{equation}\label{eq:zave}
\bar{\varepsilon}_f(x)=\frac{\lambda+1}{\lambda }\left[\left(1-F(x)\right)\nu + 1 \right] x.
\end{equation}
and 
$
\bar{\sigma}_f(x)=[1-F(x)]x.
$
Similarly, along the reverse path,
\begin{equation}\label{eq:zavereverse}
\bar{\varepsilon}_{r}(x)=\left[\left(1-F(x)\right)\nu+1 \right]x.  
\end{equation}
and 
$
\bar{\sigma}_{r}(x)=\frac{\lambda}{\lambda+1}[1-F(x)]x.
$
Both, equilibrium and out of equilibrium (averaged) stress-strain relations  are illustrated in Fig.~\ref{fig:Average_properties}. 

\paragraph{Brittle to ductile transition.}  It is   easier to see   if the system is brittle if we consider  the out-of-equilibrium (marginal stability)  path,  even though the actual ductility threshold would be the same if we consider   the global minimum path.  All we need to check is the condition that the curve  $\bar{\varepsilon}(x)$ has a local maximum, which reads
$
[1-F(x_{c})]-f(x_{c})x_{c}+\nu^{-1} =0.
$
To locate the  brittle to ductile transition we need to   find the  inflection point on the curve  $\bar{\varepsilon}(x)$ characterized by the   condition  $-2 f(x_c) -f'(x_c) x_c=0$. 

In the case of  Weibull distribution, we  obtain from the first of these two conditions
\begin{equation}
x_c= \left[\frac{1}{\rho}- W\left(-\frac{\exp{(1/\rho)}}{\rho  \nu}\right)\right]^{1/\rho },
\end{equation}
where $W(x)$ is the Lambert  function, defined through  the equation $x=W(x)e^{W(x)}$.   Then the  second condition  gives
$
\nu = e^{\frac{1}{\rho }+1}/ \rho,
$
which delineates the  boundary between  brittle and ductile regimes.

\paragraph{Statistics of avalanches.}   We first compute the avalanche distribution  for the case of the \emph{out-of-equilibrium} loading path; it will be clear that the same procedure can be   adapted for the \emph{out-of-equilibrium}  unloading path and for the reversible  \emph{equilibrium} paths.  

For an avalanche of size $s$ to take place along the  \emph{out-of-equilibrium} loading path and be associated with the failure of $k$th (in strength) spring, we must have  $\varepsilon_{k+j}\leq \varepsilon_{k},\,{\textstyle \mbox{{for}}\:}\, j=1,2,\dots, s-1$, and $\varepsilon_{k+ s}>\varepsilon_{k}$, which, following \cite{hansen_1992SM},  we called the \emph{forward condition}; to  secure that $\varepsilon_{k}$ is larger than  previous thresholds, we must also require that $\varepsilon_{j}\leq \varepsilon_{k},\,{\textstyle \mbox{{for\, all}}\:}\, j<k$, which we  call the \emph{backwards condition}. Given that  the rebinding  sequence for the unloading out-of-equilibrium path  is $\varepsilon_k^r=\frac{\lambda}{\lambda+1} \varepsilon_k$ and   for the   \emph{equilibrium} path   is   $\varepsilon_k^g=\sqrt{\frac{\lambda}{\lambda+1}}\varepsilon_k$, the avalanche condition in those two cases and the ensuing avalanche statistics will be the same as in the case of  out-of-equilibrium loading path, so it is sufficient to deal with this case only.

Since we are interested in the asymptotics for the avalanche distribution at large $N$, we assume that   $ s\ll N$. Using the ordered thresholds $\bar{x}_{i}$, and Eq.~\eqref{eq:zsup} for the sequence $\varepsilon_k$, we can then rewrite $\varepsilon_{k+j}\gtrless \varepsilon_{k} $ in the form 
\begin{equation}
\bar{x}_{k+j}\gtrless \bar{x}_{k}\left(1+\frac{j}{N-k-j+N \nu^{-1}}\right).
\end{equation}
Defining $ \displaystyle \delta_{k}= \frac{\bar{x}_{k}}{N-k+N \nu^{-1}} $, and using the assumption that  $j\ll N-k$,  we  can  simplify these  relation further 
\begin{equation}\label{eq:zpm_a}
\bar{x}_{k+j}\gtrless \bar{x}_{k}+j\delta_{k}.
\end{equation}

Note next that breaking of one spring at the elongation $\varepsilon_k$, corresponding to a threshold $\bar{x}_k=x$, raises the load on the remaining fiber by $\delta_k$. The average number of fibers that breaks as a result of this load increase is equal to the number of thresholds in the interval $(x,x+\delta_k)$, which is $N f(x) \delta_k$. Thus, the average number of fibers  breaking as a result of the failure of the $k$th  fiber  is,
\begin{equation}
g(x)=\frac{f(x) x}{1-F(x)+\nu^{-1}},
\end{equation}
where we  again used  the approximation $k/N\sim F(x)$ \cite{arnold1992firstSM}.  
For an avalanche of size $s$, the increase in load will be approximately $ s \delta_k$, which leads to  $g(x)  s$   broken  springs. The (forward) probability that the additional $ s-1$ springs  break  is then given by a Poisson distribution with the rate $g(x) s $, 
\begin{equation}
\tilde{p}_f(s,x) =\frac{(g(x) s)^{ s-1}}{( s-1)!}e^{-g(x) s}.
\end{equation}

To complete this expression, we still need to secure the  condition stating that all the $ s-1$ inequalities $\bar{x}_{k+1}<x+\delta_{k}$, $\bar{x}_{k+2}<x+2\delta_{k},\, \cdots$, $\bar{x}_{k+ s-1}<x+( s-1)\delta_{k}$ are satisfied. To this end, we divide the interval $(x,x+ s \delta_k)$ into $ s$  sub-intervals of size $\delta_k$. For our condition to be satisfied, we must have at least one threshold value   in the first interval $(x,x+\delta_{k})$, at least two in the first two intervals, and at least $ s-1$ in the first $ s-1$ intervals. To ensure  that $\varepsilon_{k+ s}>\varepsilon_k$,  there should be  no threshold values in the last interval $(x+( s-1)\delta_k,x+ s\delta_{k})$.  It can be shown that such  combinatorial problem can be solved giving   $p[ s-1, s] \sim \frac{1}{ s}$, \cite{hansen_1992SM}. We can then write the  probability that  the forward condition is satisfied in the form
\begin{equation}\label{eq:prob_fc_a}
p_f(s,x) =\tilde{p}_f(s,x) p[s-1,s] =  \frac{(g(x) s)^{ s-1}}{ s!}e^{-g(x) s}.
\end{equation}

We still need to satisfy the backward condition  that the threshold $\varepsilon_{k}$ is necessarily   bigger than its predecessors. To find the corresponding (backward) probability, we consider a finite number $n$ of such elements, $k-1, k-2,\dots,k-n$ and search for the condition  that $\bar{x}_{k-1}<x-\delta_{k}$, $\bar{x}_{k-2}<x-2\delta_{k},\, \cdots$, $\bar{x}_{k-n}<x-n\delta_{k}$.
If there are no thresholds in $(x-\delta_k,x)$, at most one in $(x-2\delta_k,x)$, at most two in $(x-3\delta_k,x),\dots$, and at most $n-1$   in  $(x-n\delta_k,x)$, then all our inequalities are fulfilled. This implies that the number $m$, not exceeding $n-1$, must be in the interval $\left(x- n\delta_{k},  x - \delta_{k} \right)$, while all the remaining $k-1-m$ thresholds   must be smaller than $x-n\delta_{k}$. The corresponding  probability is given again by a Poisson distribution,
\begin{equation}\label{eq:prob1_a}
\tilde{p}_b(s)= \frac{(g(x)n)^m}{m!}e^{-g(x)n}.
\end{equation}
 We can now  compute the probability that $m$ thresholds are randomly distributed among these $n$ intervals such that no threshold value lies in the interval $(x-\delta_k,x)$, at most one in the interval $(x-2\delta_k,x-\delta_k)$, at most two in the interval $(x-3\delta_k,x-2\delta_k)$, and so on. This is again a combinatorial problem whose  solution is   $p[m,n]\sim 1- \frac{m}{n}$ \citep{hansen_1992SM}. The probability for the backwards condition to be fulfilled is 
\begin{equation}\label{eq:bc1_a}
\begin{split}
p_b(s,x) &= \tilde{p}_b(s,x)p[m,n] \\ &=  e^{-g(x) n}\sum_{m=0}^{n-1}\frac{\left(g(x) n\right)^{m}}{m!n}\left(n-m\right).
\end{split}
\end{equation}
Rearranging the summation in Eq.~\eqref{eq:bc1_a}, we can  re-write it as,
\begin{equation}\label{eq:bc2_a}
\begin{split}
p_b(s,x)=\left(1-g\right) e^{-g(x) n}\sum_{m=0}^{n-1}\frac{\left(g(x) n\right)^{m}}{m!}\\ +e^{-g(x) n}\frac{\left(g(x) n\right)^{n}}{n!}.
\end{split}
\end{equation}
In the limit $n\rightarrow\infty$, we have  $\sum_{m=0}^{n-1}\frac{\left(g(x) n\right)^{m}}{m!}\rightarrow e^{g(x) n}$; and with the use of the Stirling approximation, $n!\approx n^n e^{-n} \sqrt{2\pi n}$, we can show that the last term in Eq.~\eqref{eq:bc2_a} vanishes for $g\leq 1$. The (backwards) probability is then,
\begin{equation}\label{eq:prob_bc_a}
p_b(s,x)= 1-g(x). 
\end{equation}

The probability of the avalanche of size $ s$ starting at the element $k$ with the threshold value $x_{k}=x$ can be now written as  the product of the forward \eqref{eq:prob_fc_a} and the backward \eqref{eq:prob_bc_a} probabilities,
\begin{equation}
p(s,x) = \frac{ s^{ s-1}}{ s!}g(x)^{ s-1} e^{- s g(x)} (1-g(x)).
\end{equation}
The final expression for the integrated avalanche distribution takes the form
\begin{equation}\label{eq:asymint}
p(s)= \frac{ s^{ s-1}}{ s!}{\displaystyle \int_{0}^{x_{c}}} \phi(x) e^{[-g(x)+\ln g(x)] s} dx,
\end{equation}
where  $\displaystyle {\phi(x)=\left[1-g(x)\right]\frac{f(x)}{g(x)}}$,  and   $x_{c}$ is the  maximum of the averaged curve $\varepsilon(x)$.

\paragraph{Asymptotic analysis.} We now focus on the tail of the distribution $p(s)$ assuming that  $N \to \infty$. 
We  use the  saddle-point approximation, which implies that the main contribution to the integral will come from the vicinity of $x=x_0$, where the  function $ h(x)=g(x)-\ln g(x)$ reaches its global minimum. To find $x_0$, we need to solve the equation
$
h'(x)=\frac{g'(x)}{g(x)}(g(x)-1)=0.
$
There are   three possibilities,
\begin{enumerate}
\item $g(x_0)\neq 1$ and $g'(x_0)= 0$ (ductile regime),
\item $g(x_0)=1$ and $g'(x_0)= 0$ (critical regime)
\item $g(x_0)=1$ and $g'(x_0)\neq 0$ (brittle regime).
\end{enumerate}

If $g(x_0)\neq 1$, and  $g'(x_0)=0$,  we can  write,
$
h(x)\approx g(x_0)-\ln g(x_0)+\frac{g''(x_0)}{2g(x_0)}(g(x_0)-1)(x-x_0)^2.
$
Then using the saddle-point approximation in \eqref{eq:asymint}, and applying the Stirling approximations $ s! \approx  s^{ s}e^{- s}\sqrt{2\pi s}$, we obtain
\begin{equation}\label{eq:assympducspin}
\begin{split}
p(s)&=\frac{ s^{ s-1}}{ s !}e^{- s h(x_0)}\phi(x_0)\sqrt{\frac{2 \pi}{ s \vert h''(x_0)\vert }}\\&\sim  s^{-2} e^{- s(h(x_0)-1)}.
\end{split}
\end{equation}

When  simultaneously $g(x_0)=1$ and $g'(x_0)=0$ we have $h''(x_0)=0$, and $h'''(x_0)=0$;  therefore  the Taylor expansion is $ h(x)\approx 1+ \frac{3 g''(x_0)^2}{4!}(x-x_0)^4$. 
We can also write $\phi(x)\approx-\frac{f(x_0) g''(x_0)}{2}(x-x_0)^2,$  which   allows us to re-write the integral \eqref{eq:asymint} in the form,
\begin{equation}
\begin{split}
p(s)&= \frac{ s^{ s-1}e^{- s}}{ s!}
\int_{0}^{x_{0}} -f(x_0)g''(x_0)(x-x_0)^2 \\
&\times e^{- s\frac{3g''(x_0)^2}{4!}(x-x_0)^4} dx.
\end{split}
\end{equation}
Computing  the integral explicitly and using Stirling's approximation we obtain 
$
p(s)\sim  s^{-9/4}.
$

In the brittle regime  we need to consider separately equilibrium and out of equilibrium paths. 

Consider first the  out-of-equilibrium path.  We need to  expand the function  $h(x)=g(x)-\ln g(x)$ up to second order to obtain
$
h(x)\approx 1+\frac{g'^{2}(x_0)}{2}(x-x_0)^{2}$. We can also expand $\phi(x)$ to  obtain $\phi(x)\approx -g'(x_0)f(x_0)(x-x_0)$. These expansions allow us  to approximate  the integral \eqref{eq:asymint} by
\begin{equation}
p(s)= \frac{ s^{ s-1}}{ s!}{\displaystyle e^{- s} \int_{0}^{x_{0}}} g'(x_0)f(x_0)(x_0-x) e^{- s\frac{g'(x)^2}{2}(x-x_0)^2} dx.
\end{equation}
Along the out-of -equilibrium path, the avalanches are counted  up to $x=x_0$; and if we  compute  the  integral explicitly, and use the Stirling approximations,  we obtain
$
p(s)\sim s^{-5/2}.
$

Consider now the equilibrium path.  The actual equilibrium SNAP event takes place at some $x_*<x_0$, given by the Maxwell construction. The counting of avalanches should be then performed only up to the point $x_*$, and in the integral \eqref{eq:asymint}, we must put  $x_{c}=x_*$. The function $h(x)$ will attain its minimum in the boundary point $x_*$, which is the upper limit of integration. In such case, the following asymptotic representation holds at $N\rightarrow\infty$   \citep{BruijnSM}
\begin{equation}
\int_{x_{inf}}^{x_{sup}}e^{-N h(x)}dx  \rightarrow \frac{e^{N h(x_{*})}}{Nh'(x_{*})}   
\end{equation} 
This allows to write,
$
p(s)\sim s^{-5/2}e^{- s(1-h(x_*))}.
$

\paragraph{Mapping on RFIM.} Using the condition $\partial_{X}\mathcal{H}=0$, we obtain $ X(\boldsymbol{x},\varepsilon)={ \frac{1}{\lambda+\Lambda}{ \left(\Lambda\varepsilon+\lambda\frac{1}{N}\sum_{i=1}^{N}x_{i}\right)}}$. If we substitute this expression back into $\mathcal{H}$ we obtain
\begin{equation}
\mathcal{H}=-\frac{1}{N^2}\sum_{i,j}Jx_i x_j-\frac{1}{N}\sum_i[Hx_i-v_i(x_i)],
\end{equation}
where $J=\frac{\lambda^2}{2(\lambda+\Lambda)}$, $H=\frac{\lambda\Lambda \varepsilon}{\lambda+\Lambda}$, and $$v_i(x_i)=u_i(x_i)+x_i^2+\frac{\lambda\Lambda \varepsilon}{2(\lambda+\Lambda)}.$$ 

\paragraph{Initial condition for the Burgers equation.}

In the case of finite $N$,  the equilibrium condition $\partial_{x_i}\mathcal{H}=0$ allows us to  write 
$$
\mathcal{H}(X,\varepsilon)=\frac{1}{N}\sum_{i=1}^N e_i(X)+\frac{\Lambda}{2}(\varepsilon-X)^2.
$$
Here, two metastable branches  $e_i=\frac{\lambda }{\lambda+1}\frac{X^2}{2}\Theta(l_i-\frac{\lambda}{\lambda+1}X)+\frac{l_i^2}{2}\Theta(X-l_i)$  are defined  in each interval $X\in [l_i,\frac{\lambda+1}{\lambda}l_i]$.  If we   choose  the branch with the minimal energy,  the remaining problem reduces to finding
 $$
\tilde{\mathcal{H}}(\varepsilon,\nu)= \min_{X\in \mathbb{R}}\left\lbrace \frac{1}{2\nu}(\varepsilon-X)^2 +q(X)\right\rbrace,
$$
where  
 $q(X)=\frac{1}{N}\sum_{i=1}^N\frac{X^2}{2}\Theta(l_i-\sqrt{\frac{\lambda}{\lambda+1}}X)+\frac{\lambda+1}{\lambda}\frac{l_i^2}{2}\Theta(X-\sqrt{\frac{\lambda}{\lambda+1}}l_i)$.  The initial data for the associated Burgers equation are  
$$\sigma_0=\partial_\varepsilon q=\frac{1}{N}\sum_{i=1}^N\varepsilon \Theta\left(l_i-\sqrt{\frac{\lambda}{\lambda+1}}\varepsilon\right).$$ 
In the limit $N\to \infty$ we have     $\frac{1}{N}\sum_{i=1}^N \Theta(l_i-\sqrt{\frac{\lambda}{\lambda+1}}X)\sim\int_{\sqrt{\frac{\lambda}{\lambda+1}}X}^{\infty}f(l)dl$   and    $\frac{1}{N}\sum_{i=1}^N \frac{l_i^2}{2}\Theta(X-\sqrt{\frac{\lambda}{\lambda+1}}l_i)\sim \int^{\sqrt{\frac{\lambda}{\lambda+1}}X}_{0}f(l)\frac{l^2}{2}dl.$  Then, in this limit, 
$$
\tilde{\mathcal{H}}(\varepsilon,\nu)=\min_{X\in \mathbb{R}}\left\lbrace \frac{1}{2\nu}(\varepsilon-X)^2 +q^\infty(X)\right\rbrace,
$$
where  $q^\infty(X)=\sqrt{\frac{\lambda}{\lambda+1}}\int_{0}^X f (\sqrt{\frac{\lambda}{\lambda+1}}X' )(X'^2/2)dX'+ 
  [1-F (\sqrt{\frac{\lambda}{\lambda+1}}X ) ](X^2/2).$ 
The initial condition for the associated Burgers equation is $ \sigma_0(\varepsilon)= \varepsilon [1-F (\sqrt{\frac{\lambda}{\lambda+1}}\varepsilon)].$ 

\end{document}